\documentclass[manuscript]{aastex631}
\usepackage{amsmath}
\allowdisplaybreaks
\usepackage{xcolor}
\usepackage{graphicx}	
\usepackage{amsmath}	
\shorttitle{Anomaly Detection of Novel Chemistry in Exoplanets}
\shortauthors{Forestano et al.}
\graphicspath{{./}{figures/}}

\begin{document}

\title{Searching for Novel Chemistry in Exoplanetary Atmospheres using Machine Learning for Anomaly Detection}

\correspondingauthor{Katia Matcheva}
\email{matcheva@ufl.edu}

\author[0000-0002-0355-2076]{Roy T.~Forestano}
\altaffiliation{Equal contribution author.}

\author[0000-0003-4182-9096]{Konstantin T.~Matchev}
\altaffiliation{Equal contribution author.}

\author[0000-0003-3074-998X]{Katia Matcheva}
\altaffiliation{Equal contribution author.}

\author[0000-0002-6683-6463]{Eyup B.~Unlu}
\altaffiliation{Equal contribution author.}

\affiliation{Institute for Fundamental Theory, Physics Department, University of Florida, Gainesville, FL 32653, USA}



\begin{abstract}
The next generation of telescopes will yield a substantial increase in the availability of high-resolution spectroscopic data for thousands of exoplanets. The sheer volume of data and number of planets to be analyzed greatly motivate the development of new, fast and efficient methods for flagging interesting planets for reobservation and detailed analysis. We advocate the application of machine learning (ML) techniques for anomaly (novelty) detection to exoplanet transit spectra, with the goal of identifying planets with unusual chemical composition and even searching for unknown biosignatures. We successfully demonstrate the feasibility of two popular anomaly detection methods (Local Outlier Factor and One Class Support Vector Machine) on a large public database of synthetic spectra. We consider several test cases, each with different levels of instrumental noise. In each case, we use ROC curves to quantify and compare the performance of the two ML techniques.
\end{abstract}





\section{Introduction}
\label{sec:introduction}

Characterization of the chemical composition of the atmospheres of extra-solar system planets is at the forefront of current exoplanetary research. The chemical makeup of a planet's atmosphere is determined by its formation; it is reshaped by its geological evolution, escape processes, interactions with the host star, its space environment; and it is potentially modified by biological activity. Therefore studying the chemical composition of a planet's atmosphere is essential not only for understanding its formation and history, but also allows us to search for tell-tell signs of presence of life.

The main observational tool for studying exoplanet atmospheres is transit spectroscopy \citep{Schneider1994,Charbonneau2000}, where a planet is observed in transmission (primary eclipse) or emission (secondary eclipse) while it passes in front or behind the host star, respectively. During a primary eclipse, a small fraction of the observed stellar flux is being absorbed or scattered by molecules or particulates in the atmosphere, which leave spectroscopic signatures in the observed spectrum. The number of available spectra of transiting planets is increasing fast with the help of ground- and space-based observations and is expected to grow dramatically with the launch of high-resolution space telescopes like JWST \citep{2016ApJ...817...17G} and dedicated exoplanet space observatories, such as the Twinkle Space Telescope \citep{2019ExA....47...29E} and the ESA Ariel mission \citep{2021arXiv210404824T}. For example, the latter is expected to observe 1000 different planets with a wide variety of parameters \citep{2019AJ....157..242E, 2022AJ....164...15E}. The sheer number of observations, coupled with the increased spectral resolution, present a computational challenge to the existing numerical tools for data analysis and retrievals of atmospheric parameters. In recent years, a number of supervised machine-learning (ML) techniques have been explored in attempts to speed up the conventional data analysis pipeline for retrieving the atmospheric chemical composition and planet parameters from observed transit spectra \citep{Waldmann2016ApJ,Marquez2018,Zingales2018,Soboczenski2018,Cobb2019,Himes2020proc,Oreshenko2020,Fisher2020,Guzman2020,Nixon2020,Himes2020,Himes2020Marge,Yip2021,Ardevol2022,Haldemann2022,Yip2022}. In this paper, we adopt an unsupervised machine learning approach, whereby, {\em without} performing an actual retrieval, we identify planets with transit spectra that are anomalous --- due to either unusual chemical composition, or inadequate assumptions in the simulation model. Unsupervised machine learning has previously been applied in spectroscopic studies of exoplanet atmospheres in order to obtain a quick preliminary classification according to chemical composition \citep{2021AJ....162..288M,2022PSJ.....3..205M} or to generate informed priors for subsequent Bayesian retrievals \citep{Hayes2020}. 

At the heart of every retrieval model that analyzes the observed spectrum is a detailed radiative transfer model (RTM) of varying complexity  \citep{Waldmann2015ApJ,Kitzmann2020,Harrington2021,Cubillos2021,Blecic2021,Welbanks2021}. The RTM takes into account the planet-star observational geometry (masses, radii, orbital parameters, etc.) and makes assumptions about the atmospheric composition and chemistry, cloud coverage and pressure-temperature profile. More advanced models include large-scale dynamical effects \citep{2023RemS...15..635P}, day/night asymmetries \citep{2022A&A...658A..42P,2022ApJ...929...20M,2022ApJ...933...79W} and latitudinal/longitudinal variations \citep{2022A&A...658A..41F}. The higher the complexity of the model, the more computationally expensive the analysis and the larger the number of ``free parameters" (and assumptions) used to fit the observations. In reality, the discovered exoplanets are very diverse, and both the number and the nature of the absorbing agents in the atmosphere are a priori unknown, and possibly quite different from our expectations. In that respect, it is rather naive to expect that the RTM is fully capturing all the relevant details of the physics and chemistry of the observed exoplanets. This motivates developing a robust, model-independent criterion to quickly flag unusual transit spectra that seem incompatible with the simulation assumptions.

In this paper we evaluate the feasibility of some popular anomaly detection machine learning techniques to identify such unusual spectra, which we refer to as ``anomalies". In general, anomaly detection methods fall into two categories: i) outlier detection, where the training data contains both normal and anomalous examples; and ii) novelty detection, where the training data is comprised of only normal examples, and the ML model does not see any anomalous examples during training. Since ideally we want to remain agnostic about the exact origin and nature of the potential anomalies, here we shall only focus on novelty detection.

The paper is organized as follows. In Section~\ref{sec:database} we introduce the database that we use for our numerical experiments. In Section~\ref{sec:feature_engineering} we describe some necessary preprocessing steps: the standardization of the data and the definition of the normal and anomalous samples to be used in the analysis. Then in Sections~\ref{sec:lof} and \ref{sec:svm} we present the results from two commonly used anomaly detection methods --- Local Outlier Factor (LOF) \citep{Breunig2000} and One-Class Support Vector Machine (1CSVM) \citep{vapnik95}, respectively. In each case, we perform four separate experiments at various levels of instrumental noise; in each experiment, a different chemical is treated as an unexpected mystery absorber, which was left out during the simulation of the training database. Section~\ref{sec:summary} is reserved for discussion and conclusions.

\section{Database Description}
\label{sec:database}

Machine-learning approaches are data-driven, hence for our purposes we require a suitable database of exoplanet transit spectra with diverse planetary demographics. Our starting point is the Ariel Big Challenge (ABC) Database permanently available at \url{https://doi.org/10.5281/zenodo.6770103} (other similar public databases include \cite{Marquez2018,Goyal2019,Goyal2020}). The ABC database was used in the 2022 Ariel Machine Learning Data Challenge \citep{Yip_competition}, which was included in the competition track at the NeurIPS 2022 conference. The goal of that challenge was to develop reliable and scalable {\em supervised} ML methods for planetary characterization. Our objective here is somewhat different --- we are interested in {\em unsupervised} ML methods for anomaly detection. Correspondingly, we shall make appropriate adaptations (additions and selections) to the original database, as described further in this section.

\subsection{Stellar and planet parameters}

The ABC dataset, described in detail in \cite{Changeat_2022}, is a synthetic spectroscopic dataset generated with the official simulators (dedicated forward RTM and instrument simulator) for the ESA Ariel mission. It contains 105,887 synthetic spectra for 5,900 unique planetary objects that were selected from the list of currently confirmed exoplanet and TESS candidates. The planet selection was made as part of the ESA Ariel Target List initiative \citep{2019AJ....157..242E, 2022AJ....164...15E}. For each planet, the stellar and planet physical parameters were fixed to their literature values, while the chemical composition of the atmosphere was randomly generated. The atmospheres were considered to be isothermal and the temperature was fixed at the equilibrium value for the respective planet. The ABC dataset underwent extensive testing and validation during the 2022 Ariel Machine Learning Data Challenge, by members of both the organizing team and the competition teams \citep{Ariel2022}. 

\begin{figure}[t]
\begin{center}
\includegraphics[width=0.9\columnwidth]{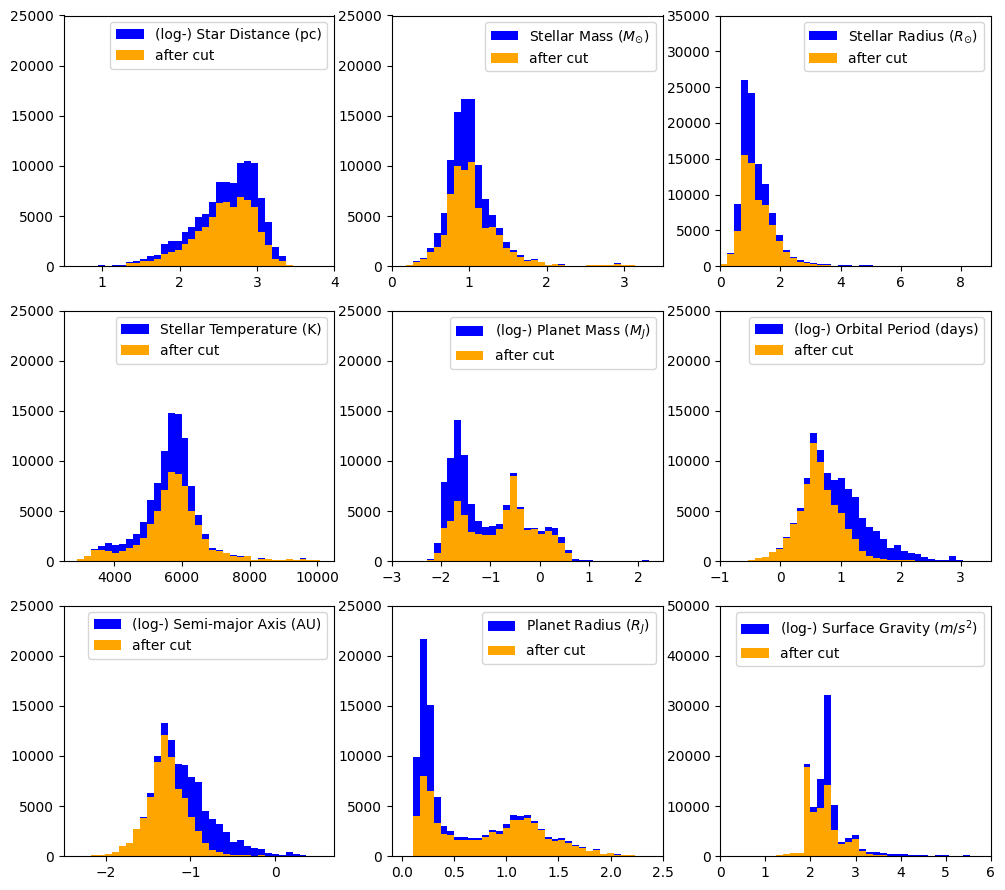}
\end{center}
    \caption{Distributions of selected stellar and planet physical parameters over the ABC dataset. Blue histograms represent the entire population in the ABC dataset, while orange histograms show the effect of the preselection cut discussed in Section~\ref{sec:preselection}.}
    \label{fig:auxiliary}
\end{figure}

\begin{figure}[t]
\begin{center}
\includegraphics[width=0.9\columnwidth]{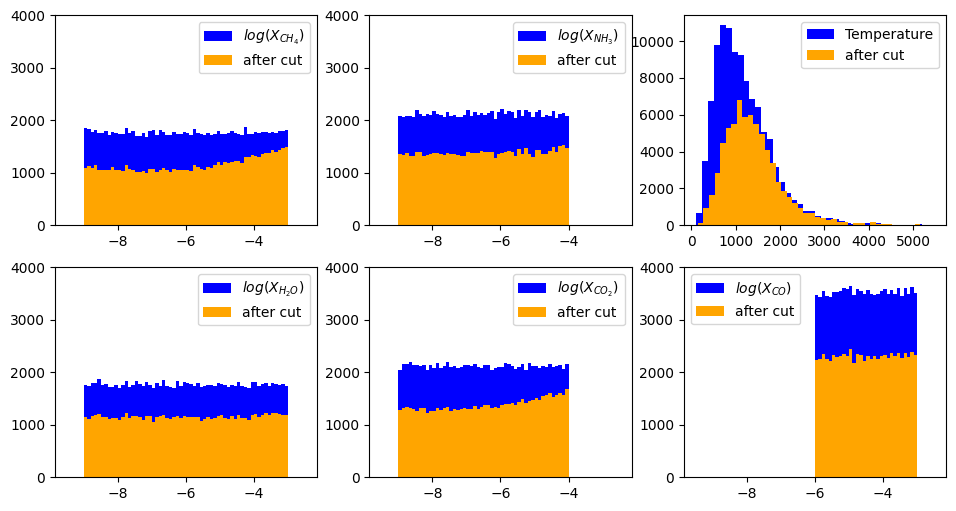}
\end{center}
\caption{The same as Figure~\ref{fig:auxiliary}, but for the planet temperature $T$ (top right panel) and the log-mixing ratios $\log(X_i)$, $i\in\{CH_4, NH_3, H_2O, CO_2, CO\}$.}
\label{fig:FMraw}
\end{figure}

The planetary population represented in the ABC dataset is quite diverse, as illustrated in Figures~\ref{fig:auxiliary} and \ref{fig:FMraw} \citep{Changeat_2022}. The blue histograms in Figure~\ref{fig:auxiliary} depict distributions of  selected stellar and planet physical parameters, including: the distance to the host star in parsecs (top left panel), the mass of the host star in units of the solar mass $M_\odot$ (top center panel), the radius $R_s$ of the host star in units of the solar radius $R_\odot$ (top right panel), the star temperature in K (middle row, left panel), the planet mass in units of the Jupiter mass $M_J$ (middle row, center panel), the planet orbital period in days (middle row, right panel), the semi-major axis of the orbit in AU (lower left panel), the planet radius in units of the Jupiter radius $R_J$ (lower center panel), and the surface gravity in SI units (lower right panel). The advantage of the ABC database is that the planets in it are potential targets for the Ariel mission. All planets are treated as hot Jupiters, with helium-to-hydrogen ratio fixed at 0.17 (which in turn also fixes the mean molecular mass).  As seen in the lower center panel in Figure~\ref{fig:auxiliary}, planets with radii below $1.5R_\oplus \approx 0.13 R_J$ were filtered out, since for them the assumption of a deep, hydrogen-dominated atmosphere is unlikely to hold. Note that the distributions in Figure~\ref{fig:auxiliary} are by no means uniform, and reflect the current observational biases --- for example, planets close to the host star (with small orbital periods) are more likely to be observed and thus represented in the dataset.

For each planet, different versions (instances) of a planet atmosphere were generated, where in addition to the primary gases ($H_2$ and $He$), each atmosphere also contains a mixture of the following five absorbers: $CH_4$, $NH_3$, $H_2O$, $CO_2$, and $CO$. Their volume mixing ratios $X_i$ were sampled on a log-uniform scale, or in other words, from a uniform distribution in terms of the log-mixing ratio $\log(X_i)$, as shown with the blue histograms in Fig.~\ref{fig:FMraw}. The sampling ranges for the mixing ratios were chosen based on the strength of the corresponding spectroscopic features and Ariel's detection capabilities  \citep{2020AJ....160...80C}:
$X_{CH_4} \in (10^{-9}, 10^{-3})$,
$X_{NH_3} \in (10^{-9}, 10^{-4})$,
$X_{H_2O} \in (10^{-9}, 10^{-3})$,
$X_{CO_2} \in (10^{-9}, 10^{-4})$,
and $X_{CO} \in (10^{-6}, 10^{-3})$.
The sampling was done independently, i.e., no particular chemical model was assumed. The $X_i$ ranges are consistent with the leading theories of planetary formation \citep{2013ApJ...763...25M,2017MNRAS.469.4102M}. The RTM model used to generate the spectra included molecular absorption, collisionally induced absorption, and Rayleigh scattering, but no haze or clouds. 

\subsection{Preselection}
\label{sec:preselection}

The synthetic benchmark dataset consists of transit spectra $M(\lambda)$ of hot Jupiters observed at 52 different wavelengths $\lambda$ in the range $0.55 \, {\mu}m - 7.275\, {\mu}m$, with variable binning according to Ariel instrument specifications. The theoretical spectra produced by the {\tt TauRex3} RTM \citep{Taurex3} were then convoluted with instrument noise simulated with the official ESA Ariel Mission radiometric model, {\tt ArielRad} \citep{ArielRad}. Unfortunately, the ideal {\tt TauRex3} spectra (before adding the noise) were not provided as part of the ABC database. Instead, we used {\tt TauRex3} to reproduce the corresponding noiseless spectra for the same values of the star and planet parameters.

In addition, the ABC dataset also includes 26,109 retrievals of approximate 6-dimensional posterior distributions (for the five absorber abundances and the temperature), using the MultiNest algorithm \citep{Multinest} as realized in {\tt TauRex3} \citep{Taurex3} and {\tt Alphnoor} \citep{2020AJ....160...80C}. Those retrievals were meant for the supervised regression-type data challenge and are not needed for the purpose of our study.

\begin{figure}[t]
\begin{center}
\includegraphics[width=0.7\columnwidth]{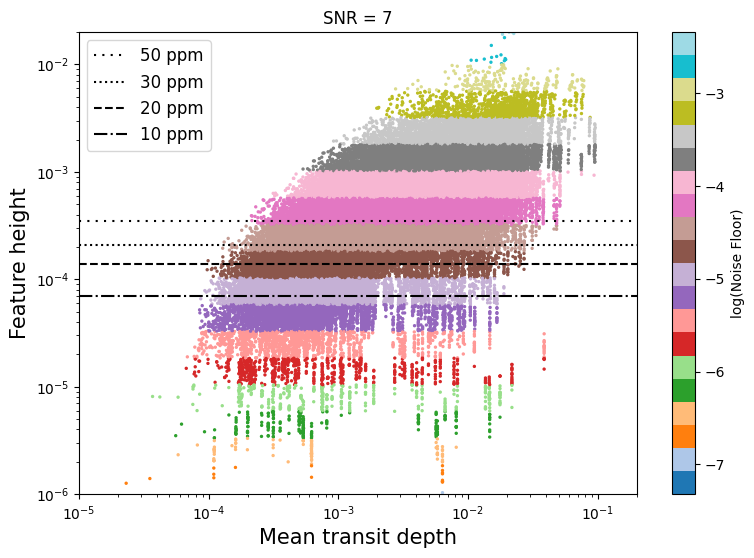}
\end{center}
    \caption{Scatter plot of all entries in the ABC database, versus the mean transit depth ($x$-axis) and the feature height ($y$-axis). Points are color-coded by the value of the noise floor $\sigma$ that would result in $SNR=7$. The horizontal lines mark the cutoff values for $\sigma=10$  ppm (dot-dashed), $\sigma=20$ ppm (dashed), $\sigma=30$ ppm (dotted), and $\sigma=50$ ppm (loosely dotted).  }
    \label{fig:cut}
\end{figure}

In order to make our analysis relevant for current and near future observations, we enforce an observability requirement for the planet spectra in the database in terms of the signal-to-noise ratio (SNR). We define the SNR as the ratio of the feature height, i.e., the difference between the maximum transit depth, $\max_\lambda(M(\lambda))$, and the minimum transit depth, $\min_\lambda(M(\lambda))$, and a Gaussian noise floor $\sigma$
\begin{equation}
SNR = \frac{\max_\lambda(M(\lambda))-\min_\lambda(M(\lambda))}{\sigma}.
\label{eq:SNR}
\end{equation}
In order to keep our analysis sufficiently general and not instrument-specific, we use a standard Gaussian noise distribution as a proxy for the instrumental noise. Our default choice for $\sigma$ is 30 ppm, but for completeness, we also show results for 10, 20 and 50 ppm. Following \cite{Changeat_2022}, we require $SNR>7$, consistent with the Ariel Tier 2 criteria. With $\sigma=30$ ppm, from (\ref{eq:SNR}) this translates into a minimum cut on the feature height of $2.1\times 10^{-4}$, as illustrated in Figure~\ref{fig:cut}. After applying this cut on the feature height, we obtain a modified database with 69,099 spectra which we use in our analysis below. The resulting parameter distributions in the modified database are depicted with orange histograms in Figures~\ref{fig:auxiliary} and \ref{fig:FMraw}. 

\begin{figure}[t]
\begin{center}
\includegraphics[width=0.85\columnwidth]{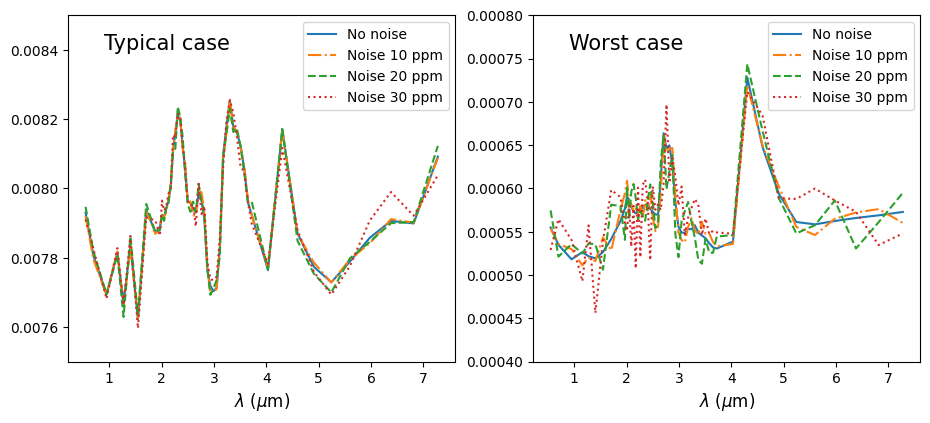}
\end{center}
    \caption{Illustration of the impact of the instrumental noise on an ideal {\tt TauRex3} spectrum (solid lines). The left panel shows a typical planet in the modified database (feature height equal to the median value of $6.3\times 10^{-4}$), while the right panel shows the planet closest to the selection cut (feature height equal to $2.100005\times 10^{-4}$). The dot-dashed, dashed and dotted lines show the effect of including instrumental noise at the level of 10 ppm, 20 ppm and 30 ppm, respectively.
    }
    \label{fig:spectra}
\end{figure}

Figure~\ref{fig:cut} shows that, even after the cut, there remains a wide variation in the feature height across the database. On one extreme, planet spectra with feature heights $\sim 10^{-2}$ have very large SNR and are hardly impacted by the instrumental noise. On the other extreme, planet spectra with feature heights near the cut of $2.1\times 10^{-4}$ are certainly feeling the impact of the noise to some extent. The worst case (closest to the cut) happens to have a feature height of $2.100005\times 10^{-4}$ and is shown in the right panel of Figure~\ref{fig:spectra}. The solid line represents the ideal {\tt TauRex3} spectrum, while the dotdashed (dashed, dotted) line includes instrumental noise at the level of 10 ppm (20 ppm, 30 ppm). A typical spectrum in the database will be somewhere between those two extremes. For example, the left panel in Figure~\ref{fig:spectra} shows the corresponding spectra (before and after inclusion of the instrumental noise) for a spectrum with feature height equal to the median value in the database, namely $6.3\times 10^{-4}$.

\section{Data Preprocessing}
\label{sec:feature_engineering}

\subsection{Definition of normal and anomalous samples}
\label{sec:normal_anomalous}

Our main goal in this paper can be formulated as follows. Given a sufficiently comprehensive database of synthetic spectra, the task is to develop a ML method which can let us quickly decide, without performing an actual retrieval, whether a given observed spectrum is unusual (anomalous) or not. An anomalous spectrum is one which is sufficiently different from all spectra in the database. In turn, this may be due to deficiencies (lack of sufficient realism and complexity) in the simulation software used to produce the database, or due to an anomalous chemical composition, e.g., an unexpected absorber which is not represented in the database. Our numerical examples will focus on the latter possibility, hence we shall define our anomalous sample to have an additional absorber which is missing in the normal population used to train the ML model. Rather than enlarging the existing ABC database with new spectra containing an additional, sixth absorber, we choose to repurpose the ABC database itself and derive both the normal and anomalous populations from it. The selection criteria are listed in Table~\ref{tab:drywet}. 
\begin{table}[t]
	\centering
 \caption{Inventory and selection criteria for the normal (columns 2 through 4) and anomalous (right three columns) examples in the database, for each of the four experiments discussed in the text. }
 \label{tab:drywet}
 \begin{tabular}{c|ccc|ccc}
 \hline
Experiment & \multicolumn{3}{c|}{Normal sample} &
\multicolumn{3}{c}{Anomalous sample} \\
  \cline{2-7}
No & Composition & Condition & Size & Mystery absorber & Condition & Size  \\
  \hline
1 & $H_2O, CO_2, CO, NH_3$   & $r_{CH_4}<10^{-4}$ & 11,169  &  $CH_4$ & $r_{CH_4}>0.05$ & 29,365 \\
2 & $H_2O, CO_2, CH_4, CO  $ & $r_{NH_3}<10^{-4}$ & 15,833  & $NH_3$  & $r_{NH_3}>0.05$ & 16,288    \\
3 & $CO_2, CH_4, CO, NH_3$   & $r_{H_2O}<10^{-4}$ & 12,106 & $H_2O$ & $r_{H_2O}>0.05$ & 26,342 \\
4 & $H_2O, CH_4, CO, NH_3$   & $r_{CO_2}<10^{-4}$ & 15,379  &  $CO_2$ & $r_{CO_2}>0.05$ & 17,869 \\
  \hline
 \end{tabular}
\end{table}
As shown in the table, we perform four separate experiments, depending on our choice of the ``mystery" anomalous ingredient. In each case, the normal population is defined as lacking that ingredient in any observable quantity. Population studies of hot Jupiters based on Bayesian retrievals have demonstrated that for the case of the five absorbers in our database, transit spectra are not sensitive to mixing ratios below $10^{-7}$, as evidenced by the flat posterior distributions \citep{Tsiaras_2018,Ariel2022}. At the same time, theoretical work has shown that the relevant quantity for detectability of an absorber $i\in \{CH_4, NH_3, H_2O, CO_2, CO\}$ is its relative abundance with respect to the other present absorbers (as opposed to relative to the hydrogen and helium, the main gases in the atmosphere). Thus we introduce the {\em relative} mixing ratio $r_i$ as the ratio of mixing ratios of the absorbers alone \citep{2022ApJ...939...95M} 
\begin{equation}
r_i \equiv \frac{X_i}{\sum_i X_i}.
\label{eq:rdef}
\end{equation}
Correspondingly, in Table~\ref{tab:drywet} planets with normal spectra are defined as having very low values of $r$ for the respective mystery absorber. Since no concentration $X_i$ was sampled above $10^{-3}$ in the creation of the database (see Figure~\ref{fig:FMraw}), a cut of $r_i<10^{-4}$ guarantees that $X_i<10^{-7}$. In contrast, an anomalous chemical composition is defined as having a mystery absorber in excess of $r=0.05$, i.e., the mystery component is at least 5\% of all absorbing agents. The described cuts leave us with four pairs of normal and anomalous populations which will be used for training and testing purposes in the examples below.

\begin{figure}[t]
\begin{center}
\includegraphics[width=0.95\columnwidth]{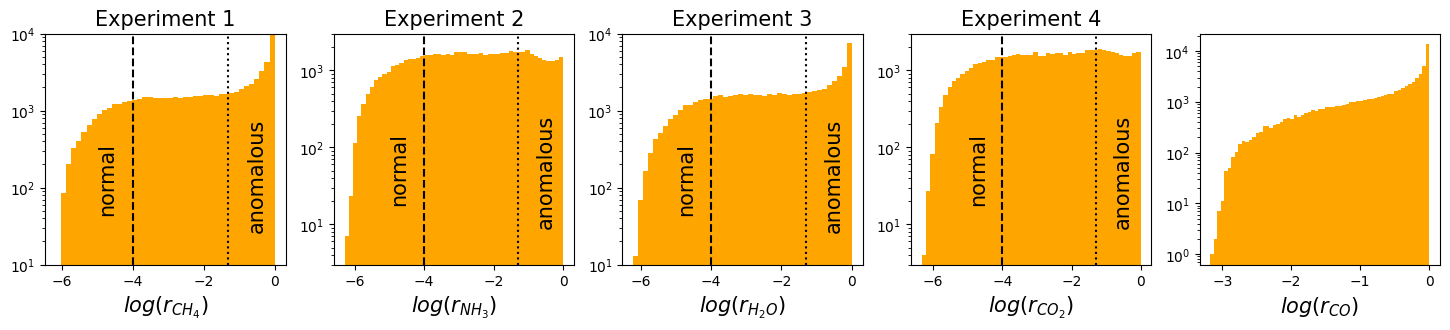}
\end{center}
    \caption{
    Distributions of the relative concentrations $r_i$, $i\in\{CH_4, NH_3, H_2O, CO_2, CO\}$, over the restricted ABC dataset.}
    \label{fig:rel_conc}
\end{figure}

\begin{figure}[t]
\begin{center}
\includegraphics[width=0.95\columnwidth]{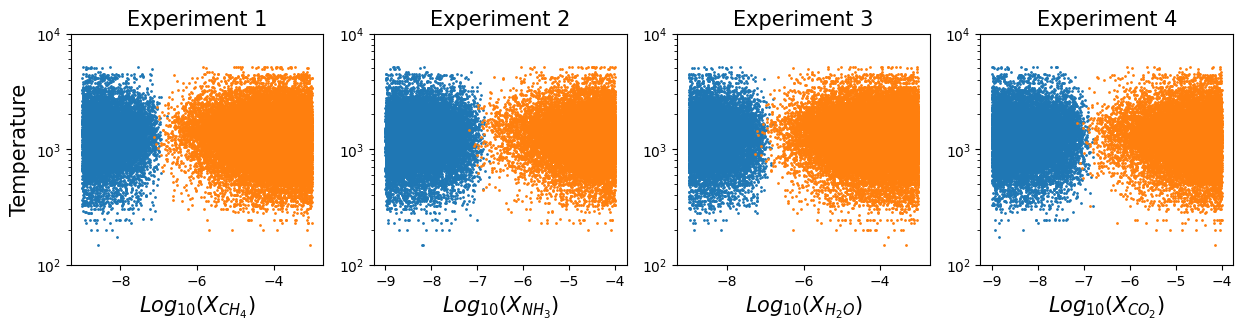}
\end{center}
    \caption{
    Scatter plots of the normal instances (blue points on the left) and the anomalous instances (orange points on the right) in the plane of the corresponding log-mixing ratio $\log(X_i)$ versus temperature $T$, for each of the four experiments discussed in the text.}
    \label{fig:clouds}
\end{figure}

The resulting normal and anomalous populations are illustrated in Figures~\ref{fig:rel_conc} and \ref{fig:clouds}. Figure~\ref{fig:rel_conc} shows the distributions of the relative concentrations $r_i$, $i\in\{CH_4, NH_3, H_2O, CO_2, CO\}$, over the modified ABC dataset of 69,099 spectra. The dashed (dotted) vertical line represents the selection cut for the normal (anomalous) population for each experiment. Note that the $r_{CO}$ distribution extends down to only about $10^{-3}$, precluding a selection of a normal sample. Therefore we do not consider an experiment in which $CO$ plays the role of a mystery absorber.

For each of the four experiments defined in Table~\ref{tab:drywet}, figure~\ref{fig:clouds} depicts the corresponding scatter plots in the $(\log X_i, T)$ plane of the normal data points (the left cluster of blue points) and the anomalous data points (the right cluster of orange points). Note that the two populations are separated around $X_i\sim10^{-7}$.   

\subsection{Standardization of the spectra}
\label{sec:preprocessing}

The standard input to any supervised machine learning algorithm is a dataset in the form of a $s\times(f+t)$ matrix containing $s$ samples of $f$ feature variables $x_i^{(j)}$ and $t$ target variables $y_i^{(k)}$:
\begin{equation}
\begin{array}{cccccccc}
    x_1^{(1)}, &  x_1^{(2)}, & \ldots , & x_1^{(f)}; & y_1^{(1)},& y_1^{(2)},& \ldots , & y_1^{(t)}\\
    x_2^{(1)}, &  x_2^{(2)}, & \ldots , & x_2^{(f)}; & y_2^{(1)},& y_2^{(2)},& \ldots , & y_2^{(t)}\\
    \vdots     & \vdots      & \vdots   & \vdots     & \vdots  & \vdots & \vdots & \vdots \\
    x_s^{(1)}, &  x_s^{(2)}, & \ldots , & x_s^{(f)}; & y_s^{(1)},& y_s^{(2)},& \ldots , & y_s^{(t)}\\
\end{array}
\label{eq:dataset}
\end{equation}
In our case, the preselection in Section~\ref{sec:preselection} resulted in $s=69,099$ spectra, each containing $f=52$ feature variables in the form of the transit depth measurements $\{M(\lambda_1), M(\lambda_2), \dots, M(\lambda_{52})\}$. The nature of the target variables, on the other hand, depends on the specific regression problem at hand. For example, in a typical inversion problem, the target variables may include the planet temperature, the planet radius, the gas mixing ratios, etc. Since our approach is unsupervised, we will not use any target variables in our analysis. Occasionally, target information will be used in the presentation of our results, to glean important insights about the method performance.

As a starting point, we find it convenient to form the features $x_i^{(j)}$ as the square root of the transit depth $M_i(\lambda_j)$ for a given planet $i$ and wavelength $\lambda_j$:
\begin{equation}
x_i^{(j)} = \sqrt{M_i(\lambda_j)}. 
\label{eq:xdef}
\end{equation}
This linearizes the relationship to the planet radius in anticipation of the linear operations which follow next \citep{Matchev2021analytical}.

Following the theoretical prescription of \cite{2022ApJ...939...95M}, we then subtract the spectral mean for each planet $i$
\begin{equation}
x_i^{(j)} ~\to~ x_i^{(j)} - \frac{1}{s}\sum_{j=1}^f x_i^{(j)}.
\label{eq:centering}
\end{equation}
This removes the contribution from the planet disk itself, as well as the average absorption in the atmosphere. The remaining signal is rescaled to a unit standard deviation
\begin{equation}
x_i^{(j)} ~\to~ \frac{x_i^{(j)} }{\sqrt{\sum_{j=1}^f \left(x_i^{(j)}\right)^2}}.
\label{eq:rescaling}
\end{equation}
As discussed in \cite{2022ApJ...939...95M}, the transformations (\ref{eq:centering}) and (\ref{eq:rescaling}) tend to remove the dependences on the various astronomical parameters like the star radius $R_s$, the planet radius $R_p$, the planet mass $M_p$, and to some extent also the planet temperature $T$. This allows us to focus on the chemical composition alone, largely removing the selection bias evident in Figure~\ref{fig:auxiliary}. Note that the standardization (\ref{eq:centering}-\ref{eq:rescaling}) differs from the usual prescription adopted in the machine learning community, where the centering (\ref{eq:centering}) and normalization (\ref{eq:rescaling}) is performed in the orthogonal direction, along the sample index $i$ instead of the feature index $(j)$.

\section{Local Outlier Factor}
\label{sec:lof}

Anomaly and outlier detection is a common unsupervised task in machine learning, where the objective is first to learn what ``normal" data looks like, and then use that knowledge to identify unusual instances (observations). As an unsupervised ML task, anomaly detection falls into the class of density estimation problems, since anomalies are expected to lie in a low density (less probable) region. Given the relevance of this problem for a wide range of applications (cyber security, financial fraud, medical diagnostics, etc.), many anomaly detection algorithms have been developed over the years, as described in the textbooks \citep{geron2017}. In agreement with the ``No Free Lunch" theorem, the performance of a given method depends on the type of dataset and values of its hyperparameters, and no single method has universal systematic advantage over the others. Therefore, the choice of method is largely a matter of preference, familiarity and availability. In this paper we use the anomaly detection tools available in {\tt scikit-learn} \citep{scikit-learn} (other libraries for anomaly detection include {\tt PyOD} \citep{zhao2019pyod}, {\tt PyNomaly} \citep{PyNomaly} and {\tt alibi-detect} \citep{alibi-detect}).

The first anomaly detection technique which we consider in this section, is the so called Local Outlier Factor (LOF) method, which compares the sample density around a given point to the density around its neighbors \citep{Breunig2000}. Note that in our case, each spectrum is a point in a 52-dimensional spectral space \citep{2022ApJ...939...95M}. In each of the four experiments, we train the model on spectra taken from the normal populations defined in Section~\ref{sec:normal_anomalous}. In each case, we do about a 4:1 train-test split of the normal population and use the larger (smaller) set for training (testing). Specifically, the training sizes in the different experiments are as follows: 9,000 in Experiment 1, 13,000 in Experiment 2, 10,000 in Experiment 3 and 13,000 in Experiment 4. We then draw a number of data points equal to the testing sample size from the anomalous populations defined in each experiment. In each of the four experiments, we consider five different exercises, each having a different noise level: ideal spectra (no noise added), 10 ppm, 20 ppm, 30 ppm and 50 ppm. For consistency, every time the training and testing is done with data having the same level of noise. The goal of all these experiments is to see i) how often anomalous atmospheres are being recognized as such (true positives), ii) what is the accompanying false positive rate for normal atmospheres being mislabelled as anomalies, and iii) what is the impact of the noise on the results.

\begin{figure}[t]
\begin{center}
\includegraphics[width=0.55\columnwidth]{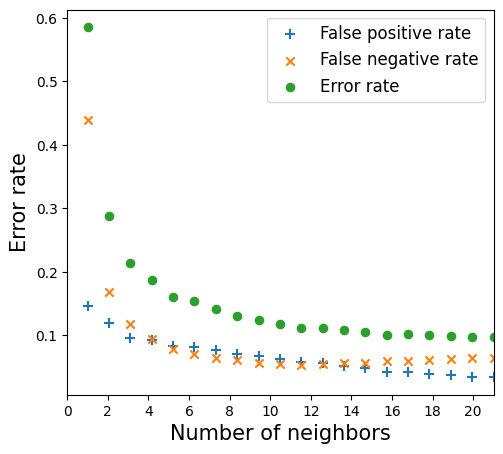}
\end{center}
\caption{Error rates as a function of the number of neighbors included in the LOF calculation. The result is for $CH_4$ (experiment 1) and noise level of 30 ppm with a cut of $-2.5$ on the LOF score.}
\label{fig:neighbors}
\end{figure}

The most important hyperparameter in the LOF method is the number of neighbors. Its default value in {\tt scikit-learn} is 20. The effect of alternative choices is explored in Figure~\ref{fig:neighbors}, where we plot the false positive rate (blue plus symbols), the false negative rate (orange crosses) and their sum (green circles) for the case of 30 ppm noise in Experiment 1. We see that the total rate plateaus above about 10 neighbors. Since the method is operating in a high-dimensional space (the 52 dimensions of the spectral features), in order to speed up the computations, we chose 10 neighbors for all results shown below.

\begin{figure}[t]
\begin{center}
\includegraphics[width=0.95\columnwidth]{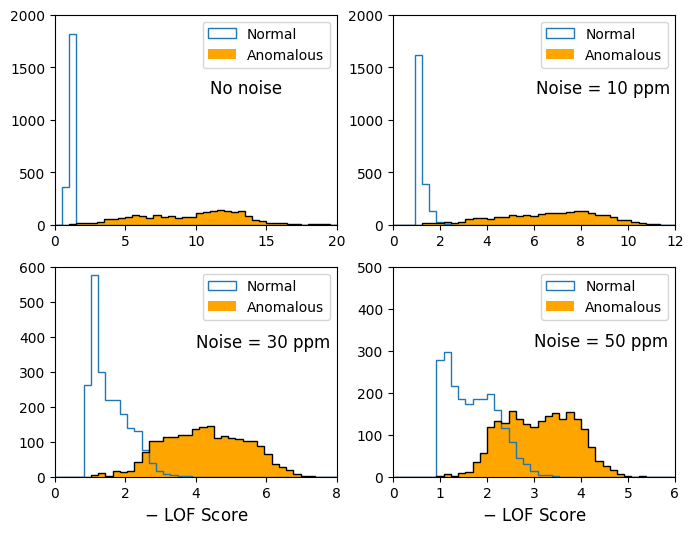}
\end{center}
    \caption{Distributions of the negative of the LOF score for the normal and anomalous populations in Experiment 1, for different levels of instrumental noise.
    }
    \label{fig:CH4_scores}
\end{figure}

Figure~\ref{fig:CH4_scores} shows one-dimensional distributions of the negative LOF score in Experiment 1 (mystery absorber $CH_4$) for four different levels of instrumental noise as labelled in each panel. The blue outlined histograms show the score distributions for the normal test data, while the orange filled histograms represent the anomalous test data. We see an excellent separation between the two test populations at the simulation truth level (upper left panel), as well as with low noise levels (10 ppm and 30 ppm). Naturally, as the noise level increases, there is more overlap between the two distributions, but they are still well separated.

\begin{figure}[t]
\begin{center}
\includegraphics[width=0.3\columnwidth]{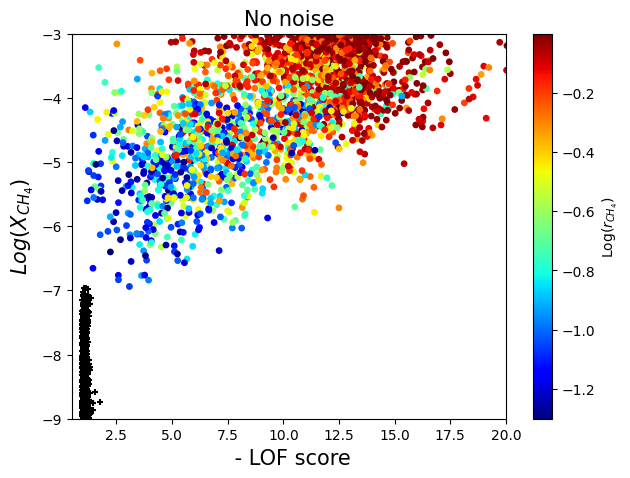}
\includegraphics[width=0.3\columnwidth]{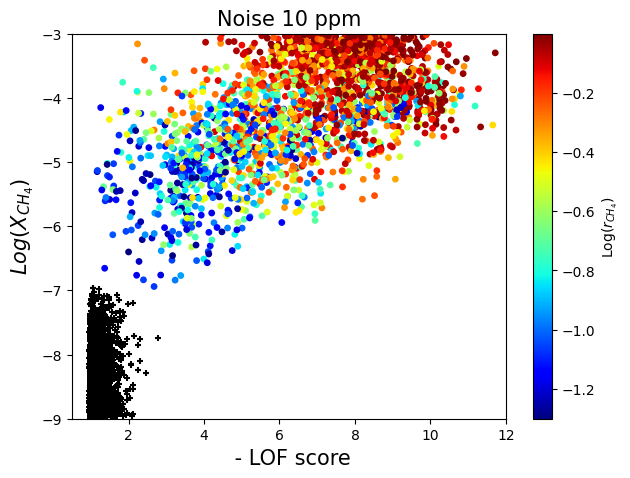}
\includegraphics[width=0.3\columnwidth]{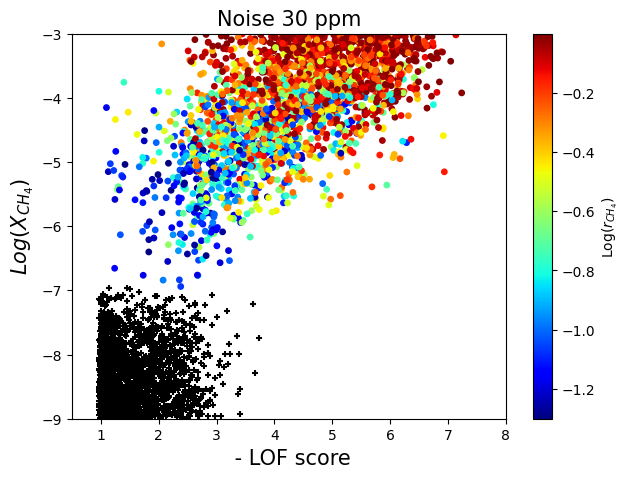}\\
\includegraphics[width=0.3\columnwidth]{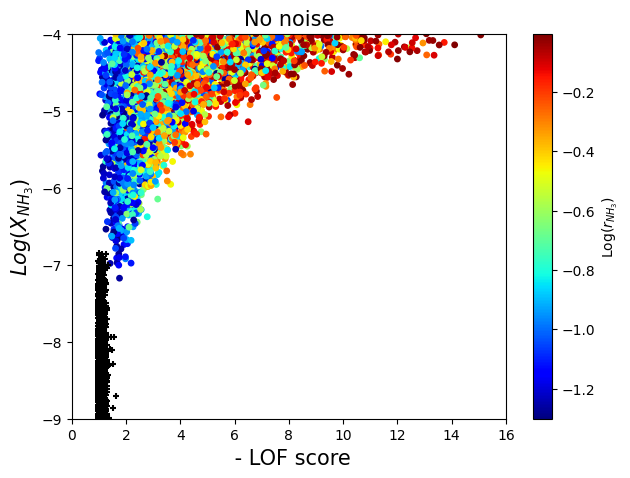}
\includegraphics[width=0.3\columnwidth]{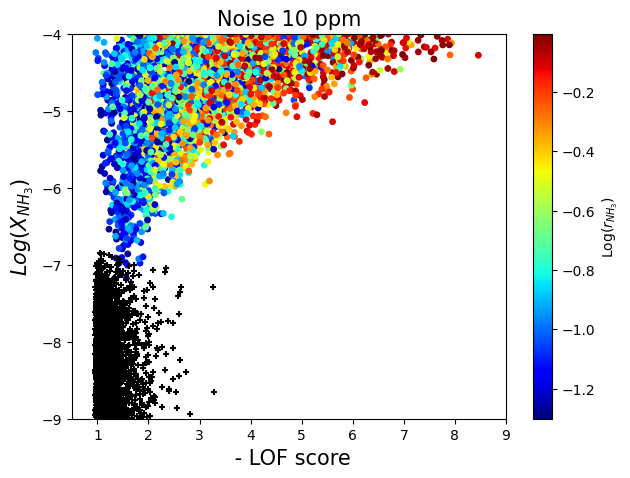}
\includegraphics[width=0.3\columnwidth]{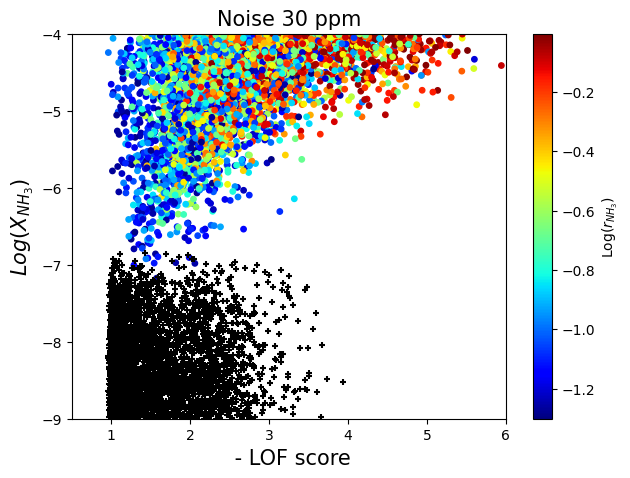}\\
\includegraphics[width=0.3\columnwidth]{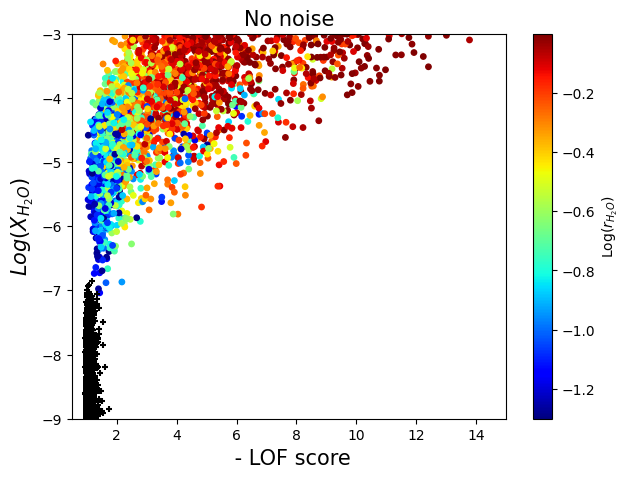}
\includegraphics[width=0.3\columnwidth]{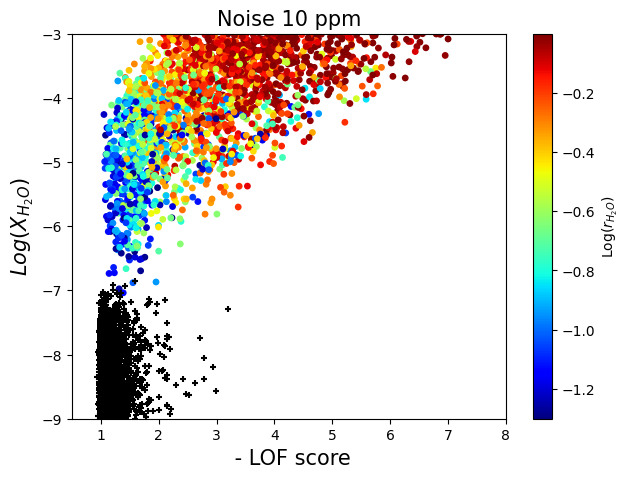}
\includegraphics[width=0.3\columnwidth]{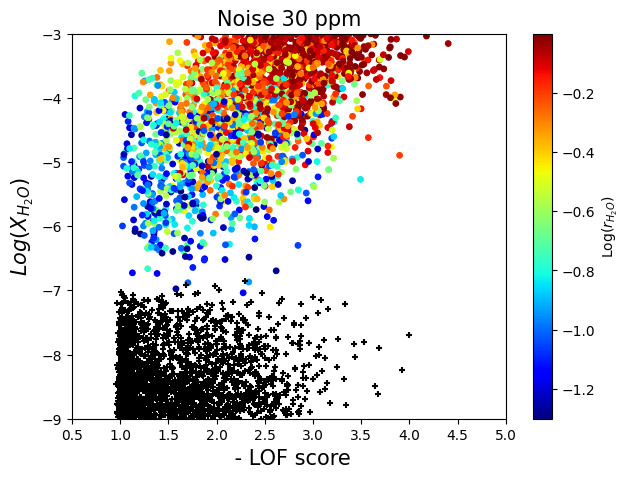}\\
\includegraphics[width=0.3\columnwidth]{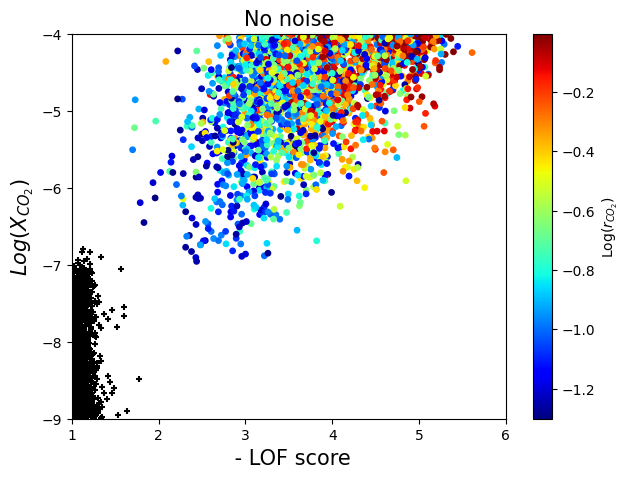}
\includegraphics[width=0.3\columnwidth]{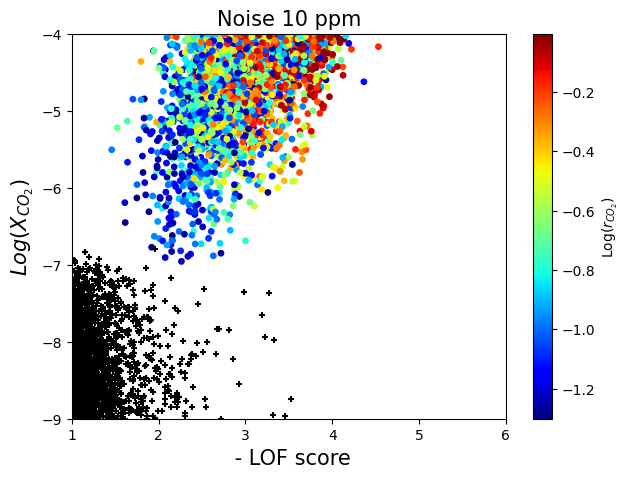}
\includegraphics[width=0.3\columnwidth]{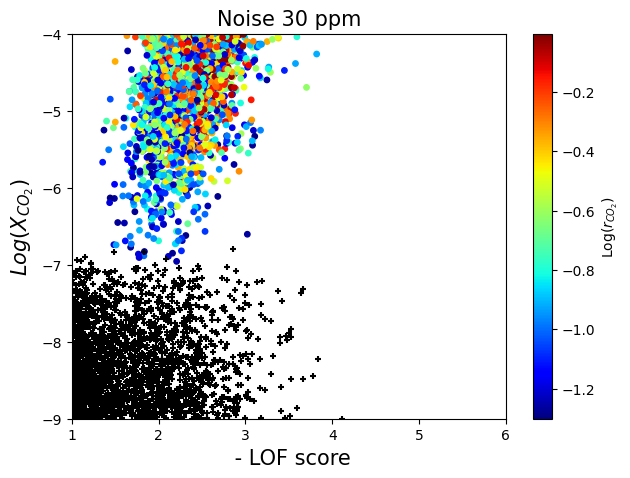}
\end{center}
    \caption{
    Scatter plot in the plane of the log-mixing ratio $\log(X_{i})$ versus the negative of the LOF score, of the normal instances (black plus symbols) and the anomalous instances (circles), color-coded by the log of the relative concentration $r_{i}$. The left panels shows ideal results with no instrumental noise, while the middle and right panels include instrumental noise of 10 ppm and 30 ppm, respectively. }
    \label{fig:LOF_scatter}
\end{figure}

Note that the score distributions for the normal test data are typically narrow. In contrast, the score distributions for the anomalous test data tend to be broader. It is therefore instructive to understand their structure in terms of the underlying planet parameters. For this purpose, in Figure~\ref{fig:LOF_scatter} we show results for all four experiments as scatter plots in the plane of the log-mixing ratio $\log(X_{i})$ versus (the negative of) the LOF score. The normal instances are denoted with black plus symbols, while the anomalous instances are shown as circles, which are color-coded according to the log of the respective relative mixing ratio parameter $r_i$. The left panels show ideal results with no instrumental noise, while the middle and right panels include instrumental noise of 10 ppm and 30 ppm, respectively.

\begin{figure}[t]
\begin{center}
\includegraphics[width=0.45\columnwidth]{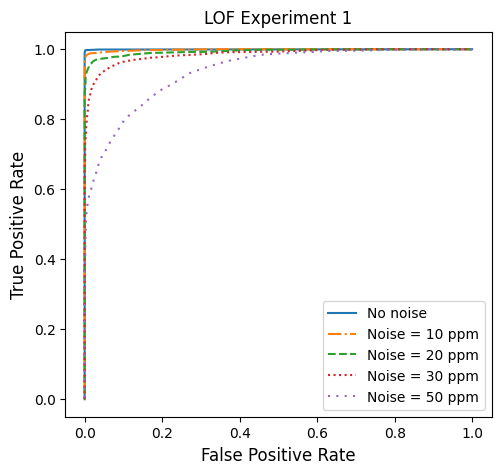}
\includegraphics[width=0.45\columnwidth]{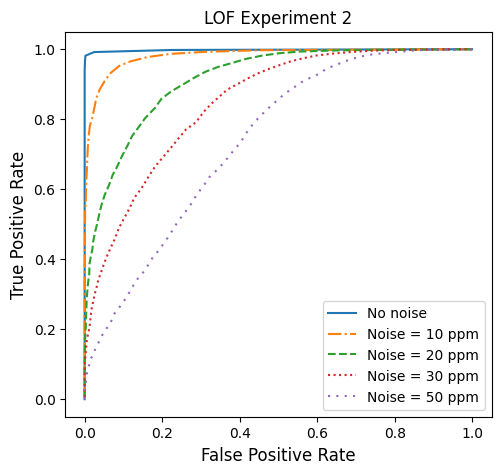}\\
\includegraphics[width=0.45\columnwidth]{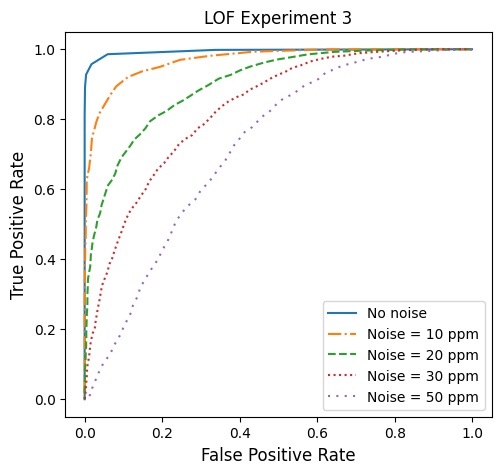}
\includegraphics[width=0.45\columnwidth]{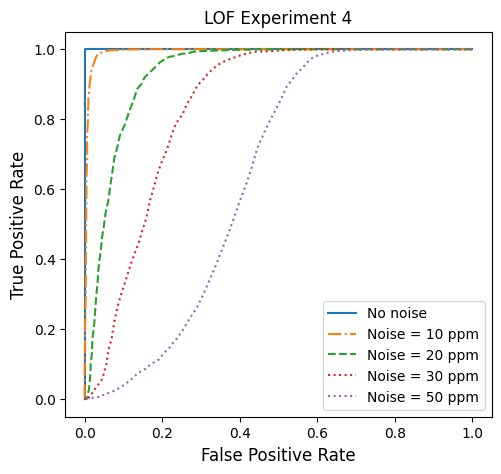}
\end{center}
    \caption{
    ROC curves obtained with the LOF method, for the four different experiments considered in Section~\ref{sec:lof}, and for different levels of instrumental noise, as described in the legends.}
    \label{fig:ROClof}
\end{figure}

While the four experiments use spectroscopically very different absorbers, the figure shows some interesting common trends. In the absence of instrumental noise, there is near perfect separation (no confusion between normal and anomalous spectra) in all experiments. As the noise level is being increased, the distributions of the normal test scores are broadening, which leads to potential confusion (false positives). The confusion rate is controlled by the separation gap between the normal and anomalous test populations, which is different for the four experiments. For example, in Experiment 1 ($CH_4$ mystery absorber), the bulk of the anomalous test population is relatively far from the normal test population, and the confusion rate stays relatively low even at the 30 ppm noise level (see the lower left panel in Fig.~\ref{fig:CH4_scores}). On the other hand, in Experiment 4 ($CO_2$) the gap is small and that results in significant confusion at larger noise levels. These observations are confirmed in Figure~\ref{fig:ROClof}, where we show the ROC curves for the four experiments. We see that the ROC curves in Experiment 1 are close to perfect, and the area under the curve (AUC) statistic is close to 1. In contrast, in Experiments 2 through 4, while the noiseless and 10 ppm ROC curves are quite good, they deteriorate more quickly with the addition of higher noise levels. 

\begin{figure}[t]
\begin{center}
\includegraphics[width=0.95\columnwidth]{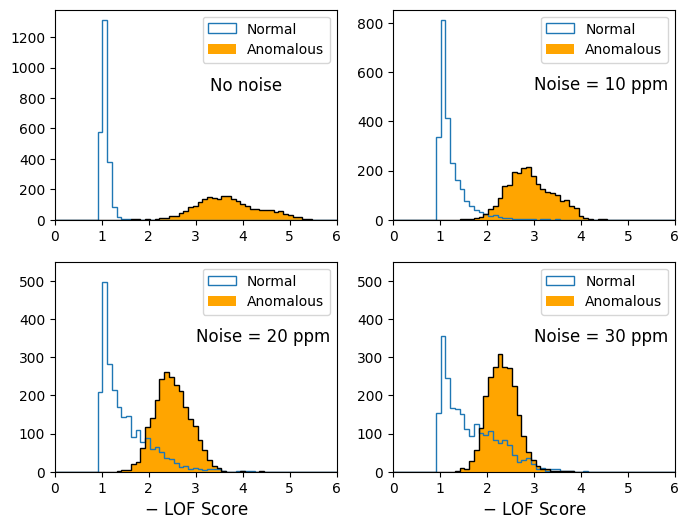}
\end{center}
    \caption{Distributions of the negative of the LOF score for the normal and anomalous populations in Experiment 4, for different levels of instrumental noise.}
    \label{fig:CO2_scores}
\end{figure}

A particularly unfortunate case is that of $CO_2$ (Experiment 4). For comparison, in Figure~\ref{fig:CO2_scores} we show the one-dimensional LOF score distributions for Experiment 4, in analogy to Figure~\ref{fig:CH4_scores}.

Coming back to Figure~\ref{fig:LOF_scatter}, the color-coding reveals that, as one might expect, the relative mixing ratio $r$ does indeed play a major role --- the larger the value of $r$, the larger the anomaly score, and hence the easier it is to tag the spectrum as anomalous. One has to appreciate that while this figure is informative, the mixing ratio $X_i$ and the relative mixing ratio $r_i$ are a priori unknown, and only the LOF score can be computed for an incoming spectral observation. Nevertheless, the LOF score alone is all that is needed to mark the planet as anomalous.

We conclude that the LOF method performs reasonably well in tagging planets with anomalous spectra for follow-up observations or additional in-depth scrutiny of their modelling. In the next section, we shall repeat our four experiments, this time using a different anomaly detection method.

\section{One Class SVM}
\label{sec:svm}

Another very popular anomaly detection ML technique is the so called One Class Support Vector Machine (1CSVM) \citep{vapnik95}. A simple linear support vector machine (SVM) is a classifier which tries to separate two classes with a plane which maximizes the width of the gap between the two classes. On the other hand, a kernelized SVM does the same, after first (implicitly) mapping the data to a higher dimensional space with the kernel trick, then fitting a linear SVM within this high-dimensional space. When applying the SVM concept to novelty detection, we have only one class --- the normal instances. The task therefore is to find a plane in the higher dimensional space which separates all training instances from the origin. In the original space this plane corresponds to a boundary which tightly surrounds the training data. If a new data point falls outside this boundary, it is treated as an anomaly.

We use the 1CSVM method available in {\tt scikit-learn} \citep{scikit-learn}. As before, the outcome depends on several  hyperparameters which should be chosen  judiciously. Two important hyperparameters are {\tt nu} and {\tt gamma}, which we select as follows: in Experiment 1 {\tt nu}=0.01 and {\tt gamma}=0.02, while in the remaining three experiments {\tt nu}=0.01 and {\tt gamma}=0.2. In all cases we use a radial basis function kernel (set by {\tt kernel}=`rbf'). The parameter {\tt nu} represents an upper bound on the fraction of training errors (i.e., training instances on the wrong side of the dividing plane). The parameter {\tt gamma} is the kernel coefficient, which controls the range of influence of each training data point on the surrounding region. Its default value is the inverse of the number of features $f$, which in our case is $1/f\approx 0.02$. In any case, hyperparameter optimization is beyond the scope of this paper, and we refer the interested reader to the specialized computer science literature on hyperparameter tuning \citep{WANG2018198}.

\begin{figure}[t]
\begin{center}
\includegraphics[width=0.3\columnwidth]{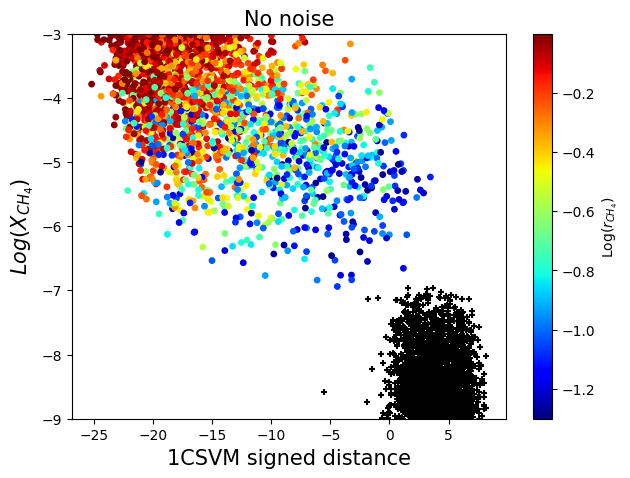}
\includegraphics[width=0.3\columnwidth]{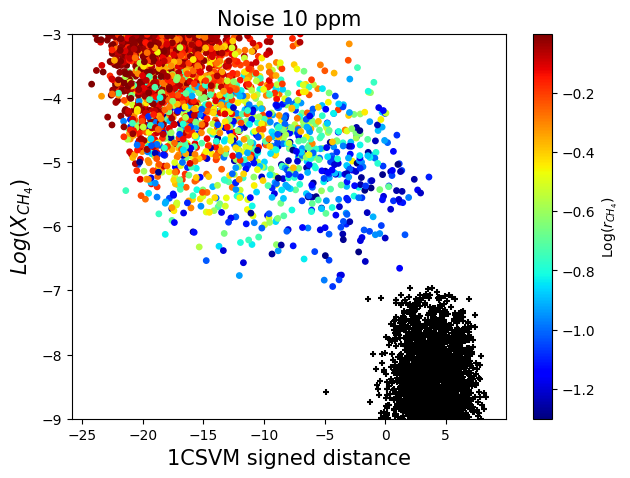}
\includegraphics[width=0.3\columnwidth]{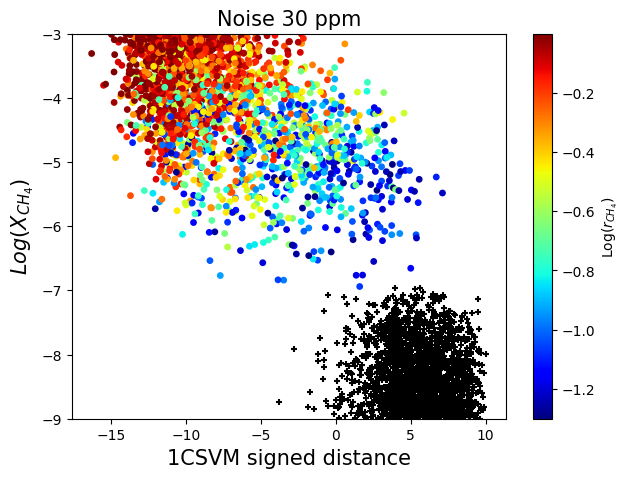}\\
\includegraphics[width=0.3\columnwidth]{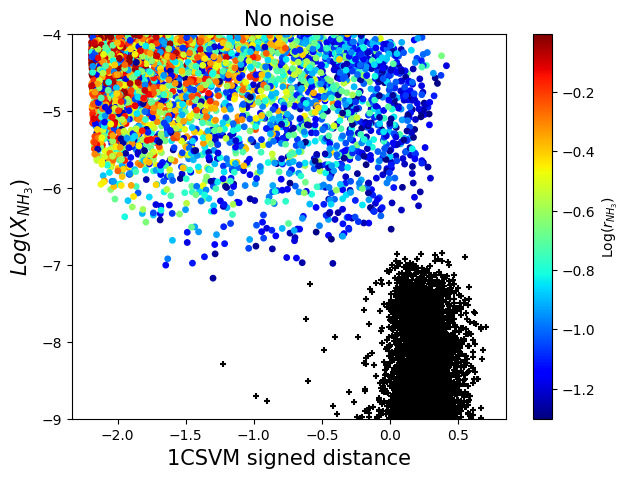}
\includegraphics[width=0.3\columnwidth]{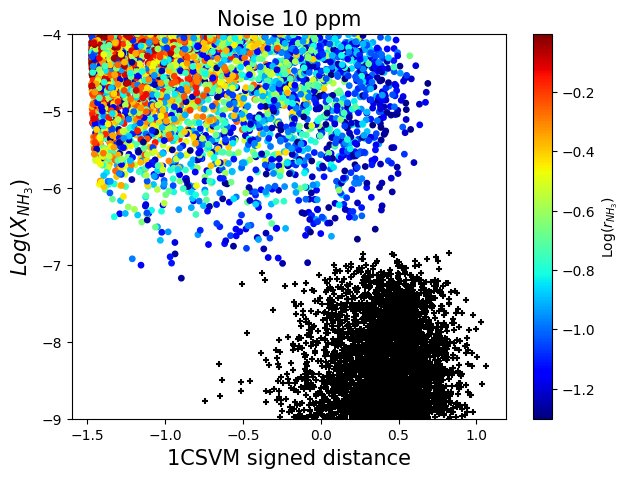}
\includegraphics[width=0.3\columnwidth]{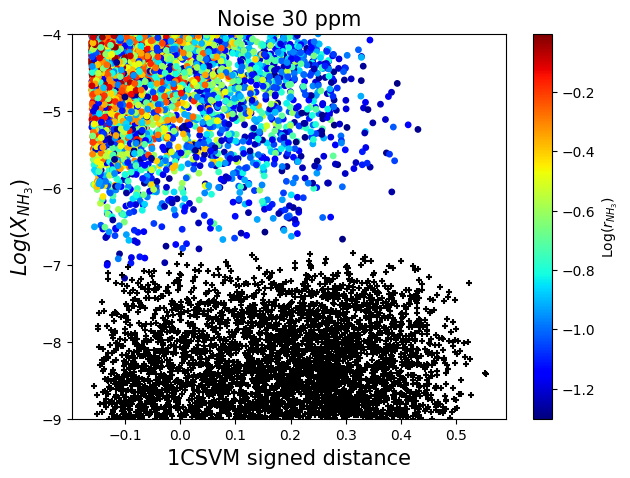}\\
\includegraphics[width=0.3\columnwidth]{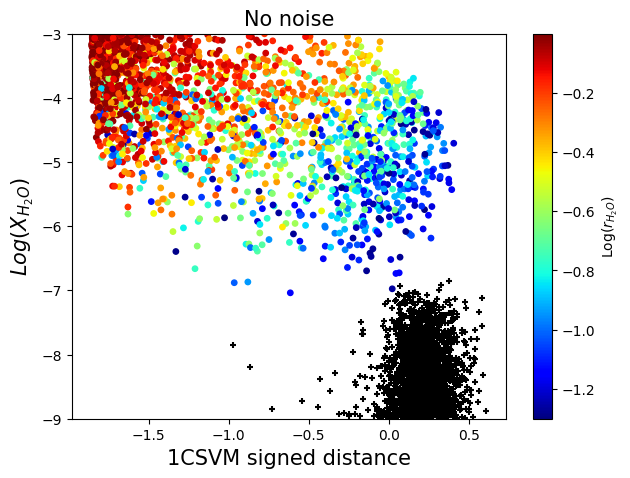}
\includegraphics[width=0.3\columnwidth]{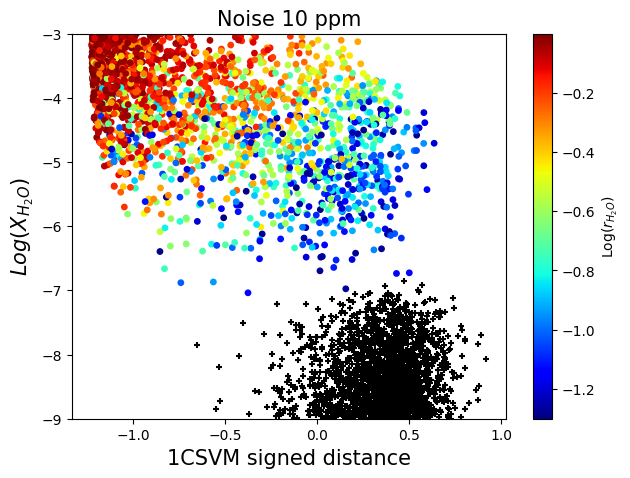}
\includegraphics[width=0.3\columnwidth]{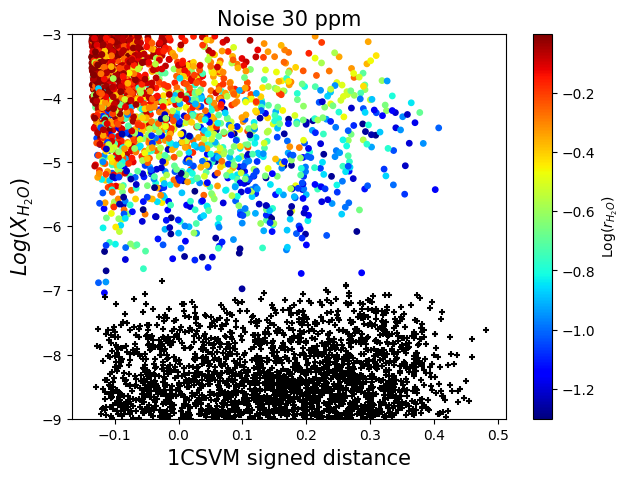}\\
\includegraphics[width=0.3\columnwidth]{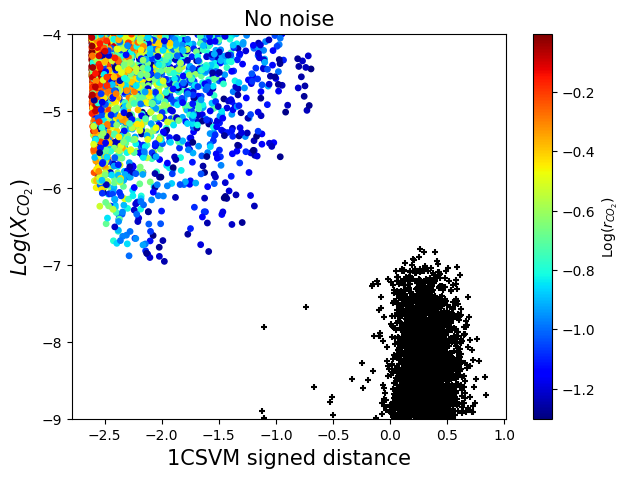}
\includegraphics[width=0.3\columnwidth]{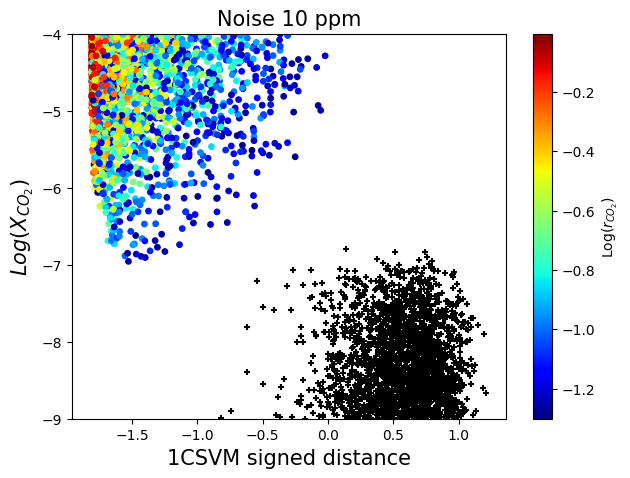}
\includegraphics[width=0.3\columnwidth]{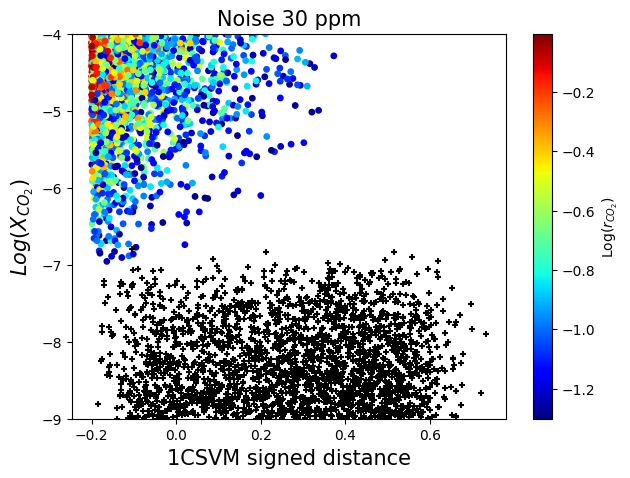}
\end{center}
\caption{The same as Fig.~\ref{fig:LOF_scatter}, but for the 1CSVM anomaly detection method discussed in Section~\ref{sec:svm}. }
\label{fig:SVM_scatter}
\end{figure}

The 1CSVM implementation in {\tt scikit-learn} returns a couple of relevant variables (scores) which are useful for our purposes. Geometrically the more intuitive is the score computed by the method {\tt \verb$decision_function()$}, which returns the signed distance to the decision boundary, with normal instances being on the `$+$' side and anomalies expected to show up on the `$-$' side. 

\begin{figure}[t]
\begin{center}
\includegraphics[width=0.4\columnwidth]{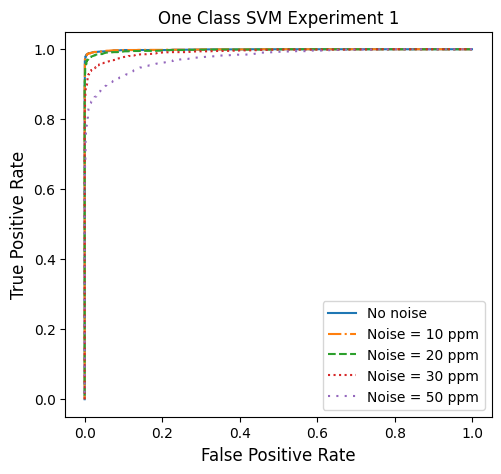}
\includegraphics[width=0.4\columnwidth]{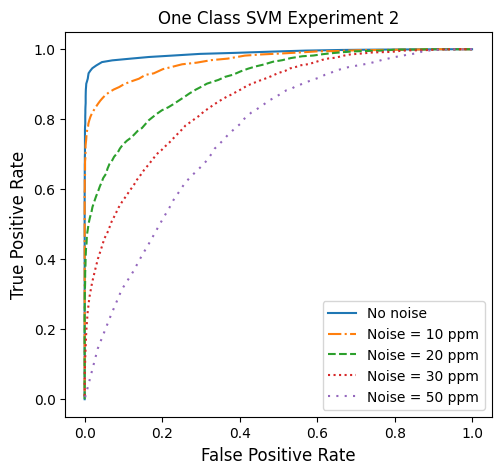}\\
\includegraphics[width=0.4\columnwidth]{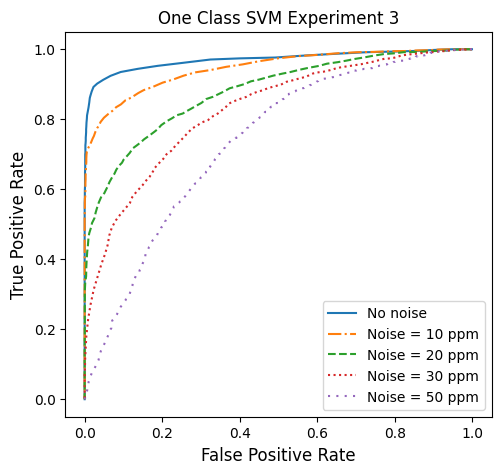}
\includegraphics[width=0.4\columnwidth]{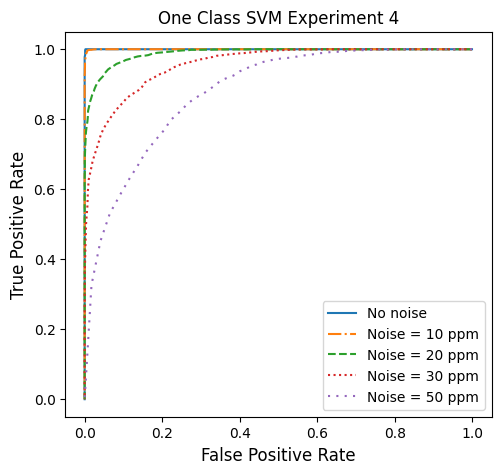}
\end{center}
    \caption{
    The same as Figure~\ref{fig:ROClof}, but for the 1CSVM anomaly detection method discussed in Section~\ref{sec:svm}.}
    \label{fig:ROCsvm}
\end{figure}

\begin{figure}[t]
\begin{center}
\includegraphics[width=0.95\columnwidth]{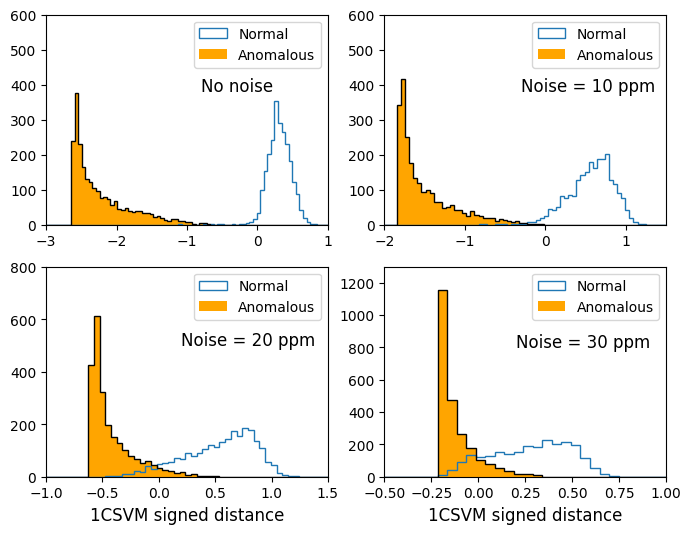}
\end{center}
    \caption{The same as Figure~\ref{fig:CO2_scores}, but for the 1CSVM anomaly detection method discussed in Section~\ref{sec:svm}.}
    \label{fig:SVM_CO2_scores}
\end{figure}

We repeat the same four experiments and show the results from the 1CSVM anomaly detection method in Figures~\ref{fig:SVM_scatter} and \ref{fig:ROCsvm}, in complete analogy to Figures~\ref{fig:LOF_scatter} and \ref{fig:ROClof}, respectively. Figure~\ref{fig:SVM_scatter} shows that the signed distance for the normal test population is largely positive, which is expected by design. At the same time, the signed distance for the anomalous population is negative and large in absolute value, and therefore the anomalous population is significantly separated from the normal population. As a result, the ROC curves in Figure~\ref{fig:ROCsvm} are very good. In fact, the ROC curves are almost ideal in the case of Experiment 1 with noise levels 10 ppm and 20 ppm, and in Experiment 4 with noise level up to 10 ppm. As the noise level increases, there is some degradation of the ROC curves, however, not as much as for the LOF method shown in Figure~\ref{fig:ROClof}. These results seem to suggest that for this particular exercise, the 1CSVM method performs better in discovering anomalies in the spectral data. To further illustrate this, in Figure~\ref{fig:SVM_CO2_scores} we plot the one-dimensional signed distance distributions in the case of Experiment 4 (mystery absorber $CO_2$), for which the LOF method was performing the worst, especially in the presence of noise. By visually comparing the horizontal separations between the normal and anomalous distributions in Figures~\ref{fig:CO2_scores} and \ref{fig:SVM_CO2_scores} one can judge the relative performance of the two methods.

\section{Discussion and Outlook}
\label{sec:summary}

The field of exoplanetary exploration has evolved significantly, transitioning from isolated planet detections to extensive surveys conducted both on the ground and in space, resulting in the identification of thousands of previously unknown planets and planetary systems. With this progression, the focus of scientific inquiry has shifted from merely detecting individual planets to conducting comprehensive planet demographics studies in search for different chemical compounds present in the atmosphere. The ultimate aim is to uncover planets with potentially habitable conditions and to discern likely indicators of biological activity. 

In this paper we advocate the use of standard machine learning techniques for the detection of anomalous chemical composition as reflected in the transit spectra observed by current and future large exoplanet surveys. We demonstrate the feasibility of two popular anomaly detection methods --- Local Outlier Factor and One Class Support Vector Machine --- on a large public database of synthetic spectra which was used in recent exoplanet data challenges. The database includes planets with a wide range of physical and chemical characteristics. Nevertheless, through an appropriate preprocessing of the spectral features we are able to eliminate the extraneous information and hone in on the chemistry-specific content of the spectrum. Our numerical experiments showed that anomalous chemical composition (here defined as the presence of an unexpected mystery absorber) is readily identifiable by those two methods.

Our study complements the work in \cite{2021AJ....162..288M}, which attempts to select interesting candidate planets for reobservation with Ariel, again without performing an atmospheric retrieval. The focus of that work was to detect a predefined molecule, by singling out the relevant wavelength bands where the corresponding spectral features are particularly strong. By comparison, our approach is completely agnostic, since we do not assume any specific knowledge about the chemistry of the observed planet, nor do we identify the reason why the spectrum appears anomalous. Other discussions of anomalous exoplanets in the literature were based on already measured chemical compositions \citep{Kinney2022} or used general planetary characteristics in lieu of spectroscopic observations \citep{2022MNRAS.510.6022S}.

Since machine learning anomaly detection methods do not pinpoint the exact reason for the anomaly, there can be several competing explanations:
\begin{itemize}
    \item {\em Discovery of an unconventional exoplanet.} This outcome stands as a particularly captivating possibility, which could indicate the presence of unexpected exotic chemistry or even physics \citep{Bai:2023mfi}, and potentially provide insights into the existence of extraterrestrial life forms.
    \item {\em Shortcomings in the data or simulations.} While perhaps lacking the immediate thrill of novel discovery, this outcome carries equal importance, pointing to deficiencies within the models employed to generate the synthetic database and define the ``normal" categories. The specific reasons could range from missing physics or chemistry to inadequate theoretical approximations or sampling ranges for the relevant parameters. At the moment, the unknown systematics present in the simulation can only be deduced by comparing the results from different codes to each other \citep{2017ApJ...850..150B,2020MNRAS.493.4884B}. 
    \item {\em Instrumental glitches and calibration errors.} Lastly, the anomalous observations could potentially stem from issues intrinsic to the observational apparatus. Such difficulties might arise from instrument malfunctions or calibration discrepancies \citep{Azari2020, Azari2021}.
\end{itemize}
These scenarios, collectively, underscore the multifaceted nature of the problem of anomaly detection. They also emphasize the intricate interplay between scientific discovery, model refinement, and observational fidelity in exoplanetary research.

\begin{acknowledgments}
We would like to thank I.~Waldmann and Gordon Yip for useful discussions. We also thank the Ariel Consortium for hospitality and financial support during the June 2023 Ariel Consortium Meeting in Spain. This work was supported in part by the United States Department of Energy under Grant No. DESC0022148.
\end{acknowledgments}


\software{
{\tt jupyter} \citep{Kluyver2016},
{\tt matplotlib} \citep{Hunter:2007ouj},
{\tt numpy} \citep{vanderWalt2011},
{\tt plotly} \citep{plotly},
{\tt scikit-learn} \citep{scikit-learn},
{\tt scipy} \citep{Scipy2020}.
}

\section*{Data Availability}

The data underlying this article are described in \citep{Changeat_2022,Yip_competition} and publicly available at \url{https://doi.org/10.5281/zenodo.6770103}.



\bibliography{references}{}
\bibliographystyle{aasjournal}

\end{document}